%% file: ms.tex
\documentclass[12pt,preprint]{aastex}

\newcommand{\teff}{$T_{\rm eff}$} 
\newcommand{\kms}{km s$^{-1}$}
\begin{document}

\title{Mg isotope ratios in giant stars of the globular clusters M 13 and 
M 71\footnote{Based on data collected at the Subaru Telescope, which is 
operated by the National Astronomical Observatory of Japan Observatory.}}

\author{David Yong}
\affil{Department of Physics \& Astronomy, University of North
Carolina, Chapel Hill, NC 27599-3255}
\email{yong@physics.unc.edu}

\author{Wako Aoki}
\affil{National Astronomical Observatory, Mitaka, 181-8588 Tokyo, Japan}
\email{aoki.wako@nao.ac.jp}

\author{David L.\ Lambert}
\affil{Department of Astronomy, University of Texas, Austin, TX 78712}
\email{dll@astro.as.utexas.edu}

\begin{abstract}
We present Mg isotope ratios in 4 red giants of the globular cluster M 13
and 1 red giant of the globular cluster M 71 based on high resolution,
high signal-to-noise ratio spectra obtained with HDS on the Subaru Telescope.
We confirm earlier results by Shetrone that for M 13, the ratio varies from 
$^{25+26}$Mg/$^{24}$Mg $\simeq$ 1 in stars with the highest Al abundance
to $^{25+26}$Mg/$^{24}$Mg $\simeq$ 0.2 in stars with the lowest Al abundance.
However, we separate the contributions of all three isotopes and 
find a considerable spread in the ratio 
$^{24}$Mg:$^{25}$Mg:$^{26}$Mg with values ranging from 48:13:39 to 78:11:11. 
As in NGC 6752, we find a positive correlation between $^{26}$Mg and Al, an
anticorrelation between $^{24}$Mg and Al, and no correlation between $^{25}$Mg
and Al. In M 71, our one star has a Mg isotope ratio 70:13:17. For both
clusters, even the lowest 
ratios $^{25}$Mg/$^{24}$Mg and $^{26}$Mg/$^{24}$Mg exceed those observed in 
field stars at the same metallicity, a result also found in NGC 6752. 
The contribution of $^{25}$Mg to the total Mg abundance is constant within a 
given cluster and between clusters with $^{25}$Mg/$^{24+25+26}$Mg $\simeq$ 0.13. 
For M 13 and NGC 6752, the ranges of the Mg isotope ratios are similar and both 
clusters show the same correlations between Al and Mg isotopes suggesting 
that the same process is responsible for the abundance variations in these 
clusters. While existing models fail to reproduce all the observed abundances, 
we continue to favor the scenario in which two generations of AGB stars
produce the observed abundances. A first generation of metal-poor AGB stars 
pollutes the entire cluster and is responsible for the large ratios of 
$^{25}$Mg/$^{24}$Mg and $^{26}$Mg/$^{24}$Mg observed in cluster stars with
compositions identical to field stars at the same metallicity. Differing
degrees of pollution by a second generation of AGB stars of the same metallicity 
as the cluster provides the star-to-star scatter in Mg isotope ratios. 
\end{abstract}

\keywords{stars: abundances --  Galaxy: abundances -- globular clusters: 
individual (M 13, M 71)}

\section{Introduction}
\label{sec:intro}

For elements heavier than Si, spectroscopic analyses of field stars and 
globular cluster stars 
have shown that they have essentially identical chemical 
abundance ratios [X/Fe] at a given metallicity, [Fe/H] (e.g., see reviews by 
\citealt{gratton04} and \citealt{sneden04}). Yet for the light elements, C-Al,
every well studied Galactic globular cluster exhibits star-to-star abundance 
variations (e.g., see review by \citealt{kraft94}). 
While the amplitude of the dispersion may differ from 
cluster to cluster, a common pattern is evident in which anticorrelations
are found between the abundances of C and N, O and Na, and Mg and Al. 
Such variations
have never been detected in field stars with comparable ages, metallicities, 
and evolutionary status \citep{pilachowski96,hanson98}. 
There is still no 
satisfactory explanation for the origin of the globular cluster abundance 
variations. 

The two explanations for the star-to-star globular cluster composition 
anomalies, the evolutionary and primordial scenarios, 
both assume that proton-capture reactions (CNO-cycle, Ne-Na chain, and 
Mg-Al chain) are responsible for altering the light element abundances. 
The principal difference is the site in which the nucleosynthesis is
thought to occur. 
In the evolutionary scenario, the observed low-mass red giants are believed 
to have altered their compositions via internal mixing and nucleosynthesis. 
To change the surface abundances of O, Na, Mg, and Al requires 
exposure to very high temperatures through very deep mixing that is
not predicted by standard models (e.g., 
\citealt{sm79}, \citealt{charbonnel95}, and \citealt{fujimoto99}). In the 
primordial scenario, the abundance anomalies are believed to be synthesized 
within intermediate-mass asymptotic giant branch (IM-AGBs) or other stars. The
present cluster members either formed from gas contaminated by these stars
or accreted such ejecta after formation. 

Measurements of the Mg isotope ratios offer a 
powerful insight into the stars responsible for the abundance variations 
because the individual isotopes are destroyed at different temperatures within
low mass red giants and IM-AGBs. 
Mg is one of the rare elements for which stellar isotope ratios can be 
accurately measured. The three stable isotopes are the alpha-nucleus $^{24}$Mg
and the neutron-rich $^{25}$Mg and $^{26}$Mg.  
Mg isotope ratios have been measured in only two globular clusters. 
\citet{shetrone96a,shetrone96b} analyzed 6 bright giants in M 13 and 
\citet{6752} (hereafter Y03) analyzed 20 bright giants in NGC 6752. 
While both analyses 
showed that the ratio $^{25+26}$Mg/$^{24}$Mg varied within each cluster, 
Shetrone was unable to separate the contribution of $^{25}$Mg from $^{26}$Mg. 
Being able to separate the contribution of each isotope is vital and only
possible when analyzing very high resolution spectra. Y03 showed
for NGC 6752 that $^{24}$Mg and Al are anticorrelated, $^{25}$Mg and Al are
not correlated, and that $^{26}$Mg and Al are correlated. Synthesis of the
large Al enhancements via proton capture on $^{24}$Mg within the Mg-Al chain 
is predicted to only occur at very high temperatures such as those found in 
IM-AGBs of the highest mass at their maximum luminosity \citep{karakas03}. 
However, the constancy of $^{25}$Mg and the correlation of $^{26}$Mg and Al 
are not predicted from IM-AGB models \citep{denissenkov03,fenner04,ventura05} 
which underscores the fact that our present theoretical knowledge of stellar 
nucleosynthesis and/or globular cluster chemical evolution is incomplete. 
Additional measurements of Mg isotope ratios in globular clusters are required 
and it is crucial to distinguish the contributions of all three 
isotopes. 

In this paper, we present measurements of the Mg isotope ratios in four bright
red giants of the globular cluster M 13 as well as one bright red giant of the 
globular cluster M 71. We re-analyze a subset of Shetrone's M 13 stars and make 
the important separation of the contributions of $^{25}$Mg from $^{26}$Mg. M 13, 
along with NGC 6752, exhibits the largest dispersion in Al abundance of all 
the well studied Galactic globular clusters. M 13 has a very similar iron
abundance to NGC 6752, [Fe/H] $\simeq$ $-$1.5. Previous studies of this cluster 
include \citet{popper47}, \citet{arp55}, \citet{helfer59}, 
\citet{cohen78}, \citet{peterson80}, 
\citet{kraft92}, \citet{shetrone96a,shetrone96b}, \citet{sneden04a}, and 
\citet{cohen05}. M~71 is a more metal-rich cluster, [Fe/H] $\simeq$ $-$0.8, 
that displays a mild O-Na anticorrelation and a subtle variation
in Al abundance. Previous studies of 
this cluster include \citet{arp71}, \citet{cohen80}, \citet{smith82}, 
\citet{leep87}, \citet{sp89}, \citet{sneden94}, \citet{ramirez02,ramirez03}, 
and \citet{lee04}. 

\section{Observations and data reduction}
\label{sec:data}

Observations of the four M 13 red giants and the M 71 red giant were obtained 
with the Subaru Telescope using the High Dispersion Spectrograph 
(HDS; \citealt{hds}) on 2004 June 1. The comparison field star, HD 141531, was
also observed since it is a red giant whose evolutionary status and stellar 
parameters are comparable to the globular 
cluster giants. Y03 measured Mg isotope ratios 
in HD 141531 allowing a comparison between data from Subaru HDS and VLT UVES. 
A 0.4\arcsec\ slit was used providing a resolving power of 
$R=\lambda/\Delta\lambda$=90,000 per 4 pixel 
resolution element with wavelength coverage from 4000~\AA~to 6700~\AA. The
integration times were about 60 minutes per star. The typical signal-to-noise 
ratio (S/N) for the program stars was 130 per pixel (260 per resolution
element) at 5140~\AA. 
One-dimensional wavelength calibrated normalized 
spectra were produced in the standard way using the IRAF\footnote{IRAF 
(Image Reduction and Analysis 
Facility) is distributed by the National Optical Astronomy 
Observatory, which is operated by the Association of Universities 
for Research in Astronomy, Inc., under contract with the National 
Science Foundation.} package of programs. In Table \ref{tab:param},
we present our list of program stars. 

\section{Analysis}
\label{sec:analysis}

\subsection{Stellar parameters and elemental abundance ratios}
\label{sec:param}

We undertook a traditional spectroscopic approach for estimating the stellar 
parameters: the effective temperature (\teff), the surface gravity (log $g$), 
and the microturbulent velocity ($\xi_t$). Equivalent widths (EWs) were 
measured for a set of Fe\,{\sc i} and Fe\,{\sc ii} lines using routines in 
IRAF where in general Gaussian profiles were fitted to the observed profiles. 
The set of Fe lines were taken from Y03, \citet{biemont91}, and 
\citet[and references therein]{blackwell95fea}. 
The EWs and atomic data are presented in Table \ref{tab:ew}. 
The model atmospheres were taken from the \citet{kurucz93} local 
thermodynamic equilibrium (LTE) stellar atmosphere grid and we used the 
LTE stellar line analysis program {\sc Moog} \citep{moog}. To set the  
effective temperature, we adjusted \teff~until there was no trend
between the abundance from Fe\,{\sc i} lines and the lower excitation potential,
i.e., excitation equilibrium. For the surface gravity, we adjusted
$\log g$ until the abundances from Fe\,{\sc i} and Fe\,{\sc ii} were equal, 
i.e., ionization equilibrium. Finally, the microturbulent velocity 
was set by the requirement that the abundances from Fe\,{\sc i} lines shows
no trend with EW. The final [Fe/H] was taken to be the mean of all Fe
lines assuming a solar abundance log $\epsilon$(Fe) = 7.50. 
In Table \ref{tab:param} we present our derived stellar parameters for
the program stars. 

For HD 141531, we compared the EWs for Fe\,{\sc i} and Fe\,{\sc ii} lines
and found an excellent agreement between the Subaru (this study) and
the VLT (Y03) measurements.
Therefore, we adopted the same stellar parameters for HD 141531 as used in Y03. 
We estimate 
the internal errors in the stellar parameters to be \teff~$\pm $ 50, 
$\log g~\pm$ 0.2, and $\xi_t~\pm$ 0.2. 
Within these uncertainties, our stellar parameters 
for the M 13 giants compare very well with those
derived by \citet{shetrone96a,shetrone96b}, \citet{sneden04a}, and 
\citet{cohen05}. For
M 71 A4, our stellar parameters agree very well with those derived
by \citet{sneden94} and \citet{shetrone96a,shetrone96b}. (See Table 
\ref{tab:comp} for comparison of stellar parameters.) 

We also measured abundances for O, Na, Mg, and Al using the same lines
presented in Y03 (atomic data and EWs are presented in 
Table \ref{tab:ew}). In Table \ref{tab:abund} we present the abundance
ratios [Fe/H] and [X/Fe] for the program stars assuming solar abudances 
of 8.69, 6.33, 7.58, and 6.47 for O, Na, Mg, and Al respectively. 
In this table, we also
compare our abundance ratios with previous studies of these stars, 
namely \citet{sneden94,sneden04a}, \citet{shetrone96a,shetrone96b}, 
and \citet{cohen05}. 
In general, the abundance ratios are in good agreement with 
previous determinations. However, for [O/Fe] our two measurements do
not agree with those derived by \citet{sneden04a}. (We note that they
reference the O abundances to Fe\,{\sc ii} and that in some cases there 
is a difference between the abundances from neutral and ionized iron.) 
To ensure consistency 
with their previous studies on globular clusters, 
\citet{sneden04a} adopt the ``traditional'' solar oxygen abundance 
log $\epsilon$(O) = 8.93 \citep{anders89} rather than the 
revised value of 8.69 recommended by \citet{allendeO}. The adopted 
solar oxygen abundance accounts for the offset between our abundances
and those of \citet{sneden04a}. Similarly, Shetrone used the traditional 
solar oxygen abundance which explains the difference in derived abundances. 
\citet{cohen05} find different abundances
for M 13 L70 (II-67) which cannot be attributed to the adopted solar 
abundances or stellar parameters. We note that they find a 0.3 dex discrepancy
between Fe\,{\sc i} and Fe\,{\sc ii}. 
In Table \ref{tab:parvar}, we present the
abundance dependences upon the model parameters. 

\subsection{Mg isotope ratios}

\begin{figure}
\epsscale{0.8}
\plotone{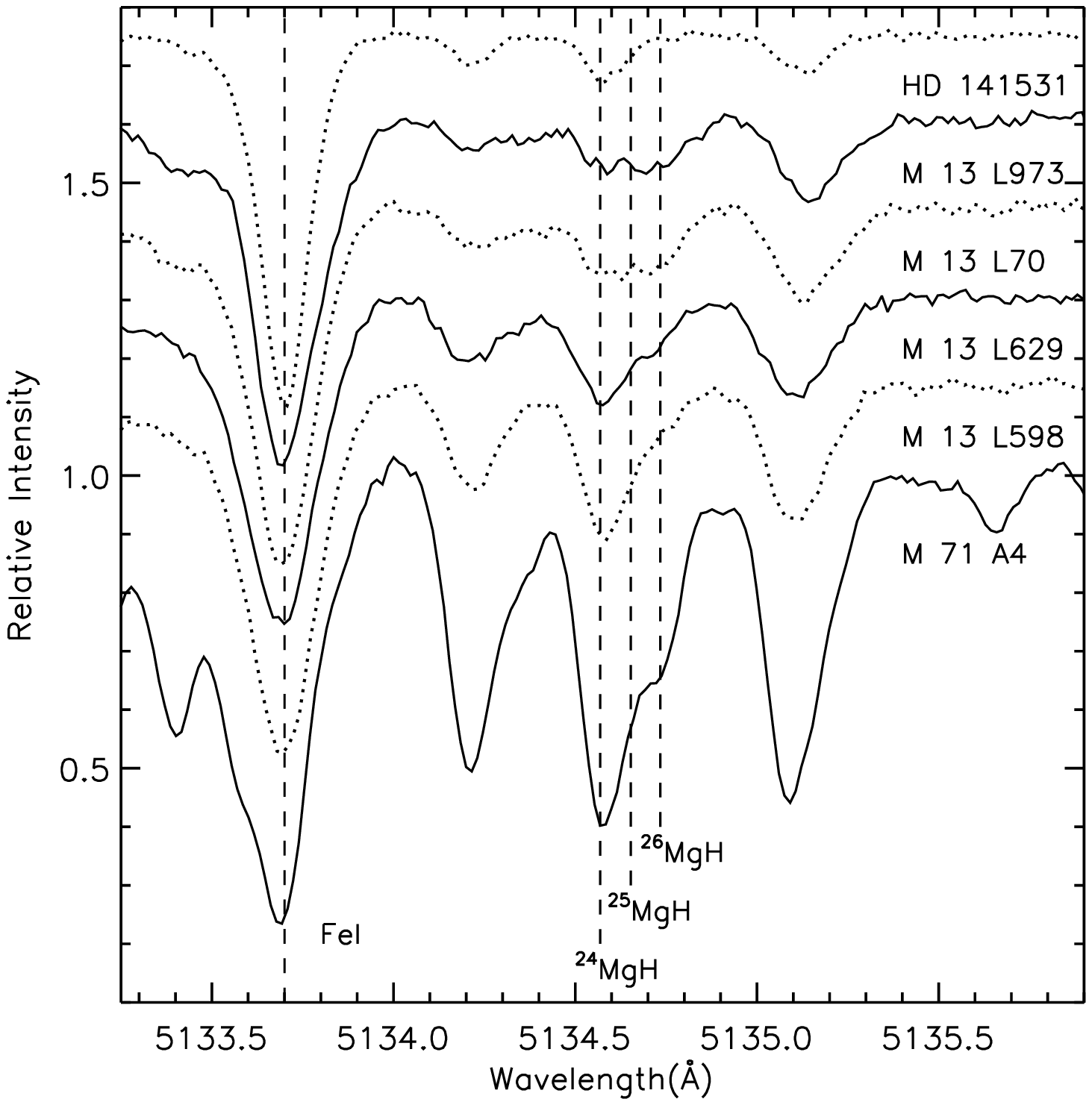}
\caption{Spectra for all program stars between 5133~\AA~and 5136~\AA. This
region includes the 5134.6~\AA~MgH line. The positions of the $^{24}$MgH, 
$^{25}$MgH, and $^{26}$MgH lines are marked by a vertical line. This MgH 
line shows substantial asymmetries for 
M 71 A4, M 13 L70, and M 13 L973. A vertical line shows a strong 
Fe\,{\sc i} at 5133.7~\AA~whose profile is generally symmetric. \label{fig:MgH}}
\end{figure}

The first measurements of stellar Mg isotope ratios were performed by 
\citet{boesgaard68} who examined the 0-0 band of the A-X electronic
transition of the MgH molecule. 
As a result of the isotopic wavelength splitting, 
$^{25}$MgH and $^{26}$MgH contribute a red 
asymmetry to the main $^{24}$MgH line (see Figure \ref{fig:MgH}). Subsequent 
studies have also exploited the molecular MgH lines to measure Mg isotope 
ratios in cool dwarfs and giants (e.g., \citealt{tl76,tl79,tl80}, 
\citealt{barbuy85,barbuy87}, \citealt{bpp87}, \citealt{lm86}, \citealt{ml88}, 
\citealt{gl2000}, \citealt{mghdwarf,mghhyades} and references therein). 
Pre-solar grains allow measurements of Mg isotope ratios with exquisite 
precision and offer a powerful additional insight into stellar nucleosynthesis
(e.g., \citealt{nittler03} and \citealt{clayton04}).

Inspection of Figure \ref{fig:MgH} shows that the profiles of the MgH lines
are asymmetric with $^{25}$MgH and $^{26}$MgH providing red wings. This
figure also shows that the profiles of the MgH lines differ significantly 
among the M 13 giants. 
In particular, M 13 L973 and M 13 L70 appear to have substantial contributions
of $^{26}$MgH relative to M 13 L 629 and M 13 L 598. M 71 A4 also exhibits 
an asymmetric profile suggesting a large contribution from $^{26}$MgH. 
While we can gain a qualitative appreciation that the Mg isotope ratios 
vary within M 13 by looking at the spectra, spectrum synthesis is 
necessary to measure accurate ratios. 
Next we describe our approach which is identical to \citet{gl2000}, Y03, and 
\citet{mghdwarf,mghhyades}. 

Numerous lines from the A-X electronic transition of 
the MgH molecule are present in the spectra of these
cool red giants. However, few of these lines offer a reliable 
measure of the Mg isotope ratios due to the presence
of known and unknown blends. \citet{ml88} recommended 3 lines from
which accurate isotope ratios can be extracted. The first line at 
5134.6~\AA~is due to the $Q_1(23)$ and $R_2(11)$ lines from the 0-0 band. The 
weaker
MgH features on either side of the 5134.6~\AA~line are contaminated such that
reliable isotope ratios cannot be measured \citep{tl80}. The second line at
5138.7~\AA~is a blend of the 0-0 $Q_1(22)$ and 1-1 $Q_2(14)$ MgH lines. The 
third line at 5140.2~\AA~is a blend of the 0-0 $R_1(10)$ and 1-1 $R_2(4)$ MgH 
lines. We refer to these lines as Region 1, Region 2, and Region 3 
respectively. 

To measure the Mg isotope ratios, we generated synthetic spectra using 
{\sc Moog}. The macroturbulent broadening was assumed to have a Gaussian form 
and was estimated by fitting the
profiles of the Ni\,{\sc i} line at 5115.4~\AA~and 
the Ti\,{\scshape i} line at 5145.5~\AA. These clean atomic lines 
were slightly stronger than the three recommended MgH lines. 
These lines gave the same macroturbulence within 1.0 km s$^{-1}$ and the 
smaller value was adopted if there was a disagreement. Our
list of atomic and molecular lines was identical to the \citet{gl2000} list
and includes the contributions from C, Mg, Sc, Ti, Cr,
Fe, Co, Ni, and Y. For the MgH isotopic components, the wavelengths were taken 
from \citet{ml88} and were based on direct measurements of an MgH spectrum 
obtained using a Fourier transform spectrometer by \citet{bernath85}. 

\begin{figure}
\epsscale{0.8}
\plotone{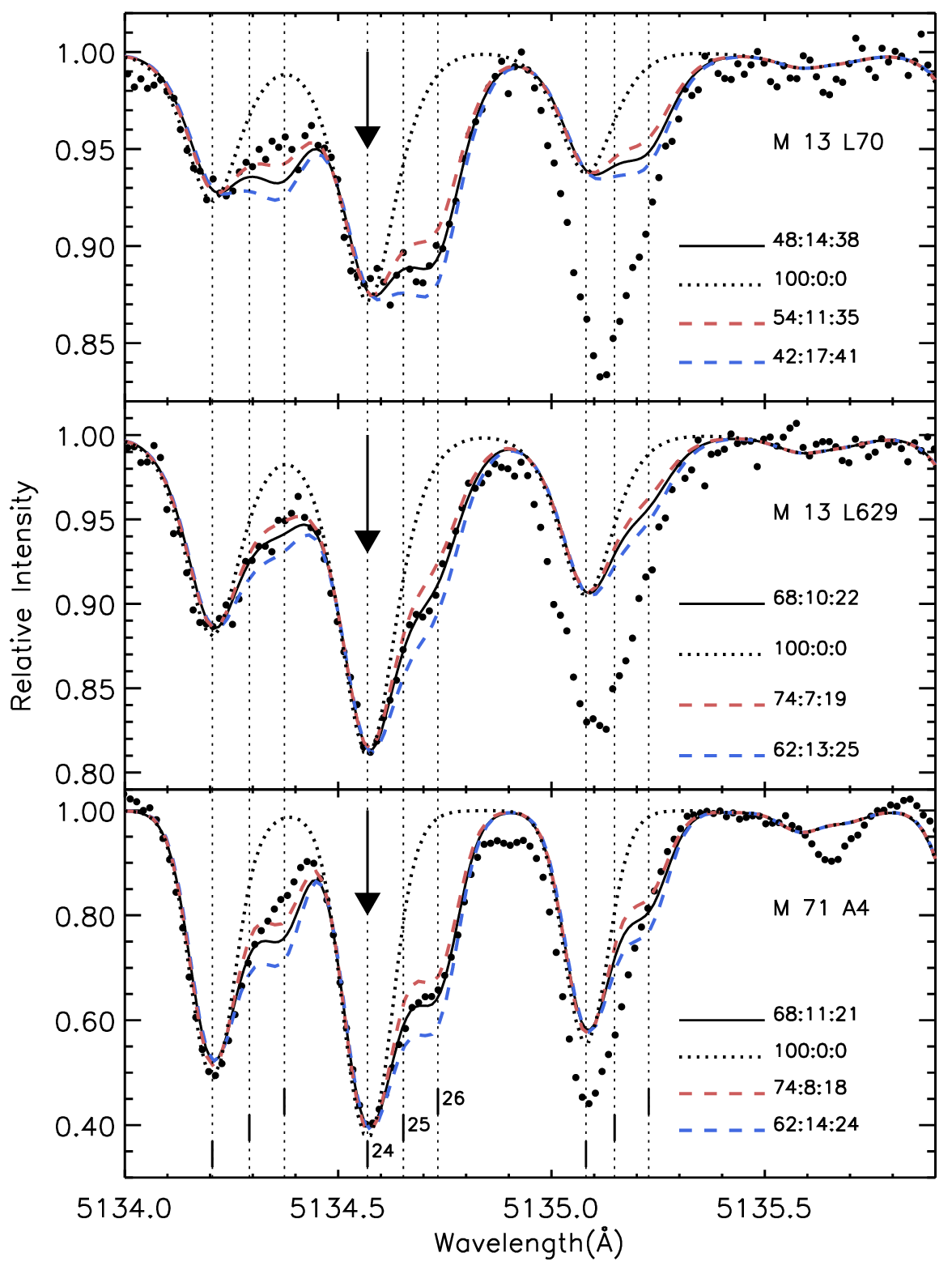}
\caption{Spectra of M 13 L70, M 13 L629, and M 71 A4 for Region 1. 
The feature we are
fitting is highlighted by the black arrow. The positions of $^{24}$MgH,
$^{25}$MgH, and $^{26}$MgH are marked by dashed lines. Filled circles
represent the observed spectra, the best fit is the solid line, and
unsatisfactory ratios are also shown. \label{fig:region1}}
\end{figure}

\begin{figure}
\epsscale{0.8}
\plotone{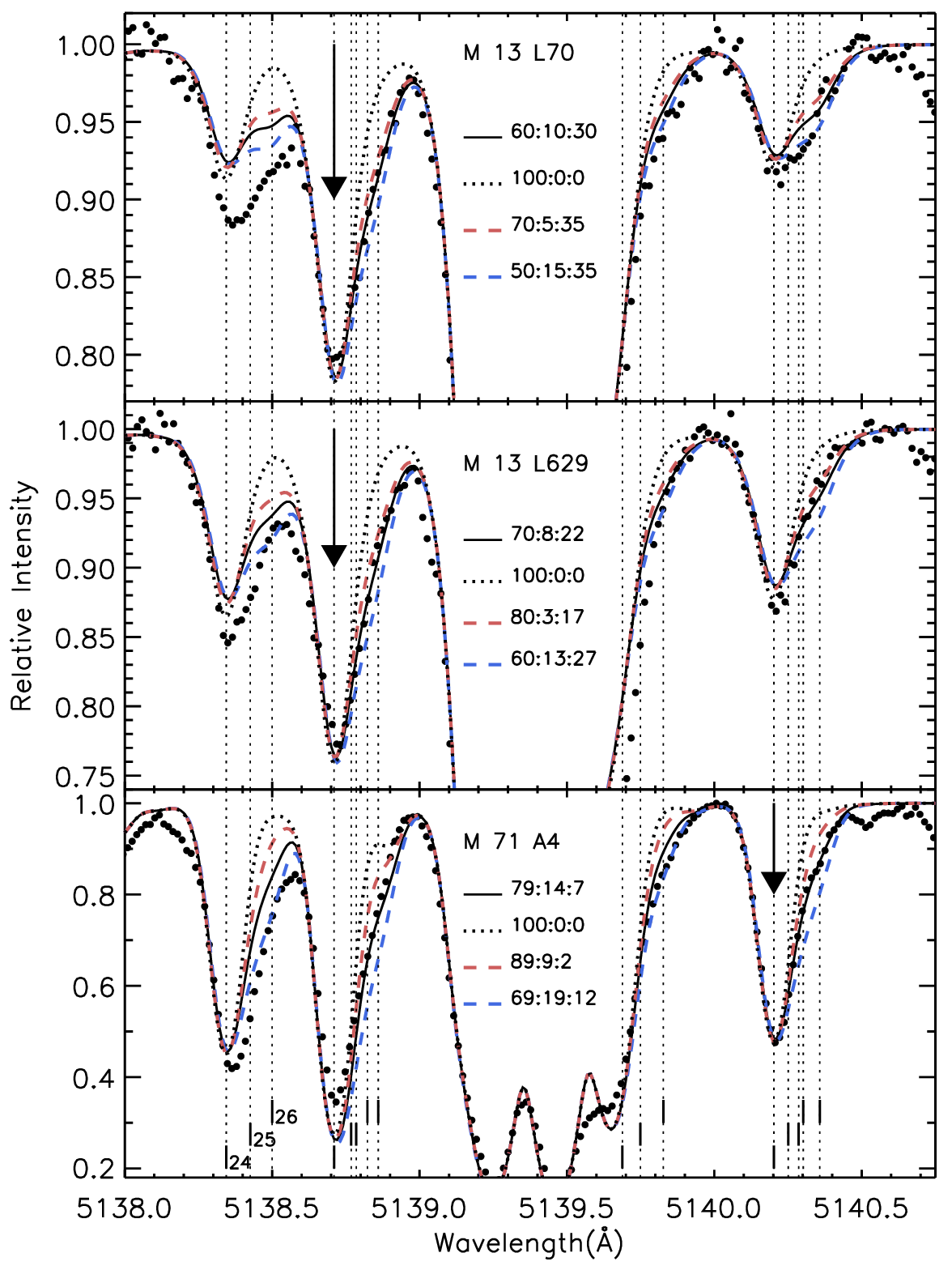}
\caption{Spectra of M 13 L70, M 13 L629, and M 71 A4 for Regions 2 and 3. 
The feature we are
fitting is highlighted by the black arrow. The positions of $^{24}$MgH,
$^{25}$MgH, and $^{26}$MgH are marked by dashed lines. Filled circles
represent the observed spectra, the best fit is the solid line, and
unsatisfactory ratios are also shown. \label{fig:region23}}
\end{figure}

Treating each recommended line independently, we adjusted the isotope ratios 
and Mg abundance until the profile was best fit. 
Following \citet{nissen99,nissen00} who measured Li isotope ratios, we chose
to use a $\chi^2$ analysis to determine the best fit to the data. The advantages
are that this method is unbiased and allows us to quantify the errors to the
fits. The free parameters were (1) $^{25}$Mg/$^{24}$Mg, (2) $^{26}$Mg/$^{24}$Mg, 
and (3) the Mg abundance, log $\epsilon$(Mg). Our initial guess was the best
fit as determined by eye. We explored a large volume of parameter space around 
the initial estimate and calculated $\chi^2 = \Sigma(O_i-S_i)^2/\sigma^2$ where 
$O_i$ is the observed spectrum point, $S_i$ is the synthesis, and 
$\sigma = (S/N)^{-1}$. Optimum values were determined by locating the 
$\chi^2$ minima for each free parameter 
$^{25}$Mg/$^{24}$Mg, $^{26}$Mg/$^{24}$Mg, and 
log $\epsilon$(Mg). We then used a more refined grid and searched a 
smaller volume in parameter space centered upon the optimum value. 
For each region in each star, we generated and tested over 500 synthetic
spectra. Throughout the various iterations for each region in each star, 
the optimal isotope ratio always 
converged to a similar value. The best fit as determined by eye was always 
similar to the best fit as determined via the $\chi^2$ analysis. 
In Figures \ref{fig:region1} and \ref{fig:region23}, we 
show examples of the best fitting synthetic spectra. 

\begin{figure}
\epsscale{0.8}
\plotone{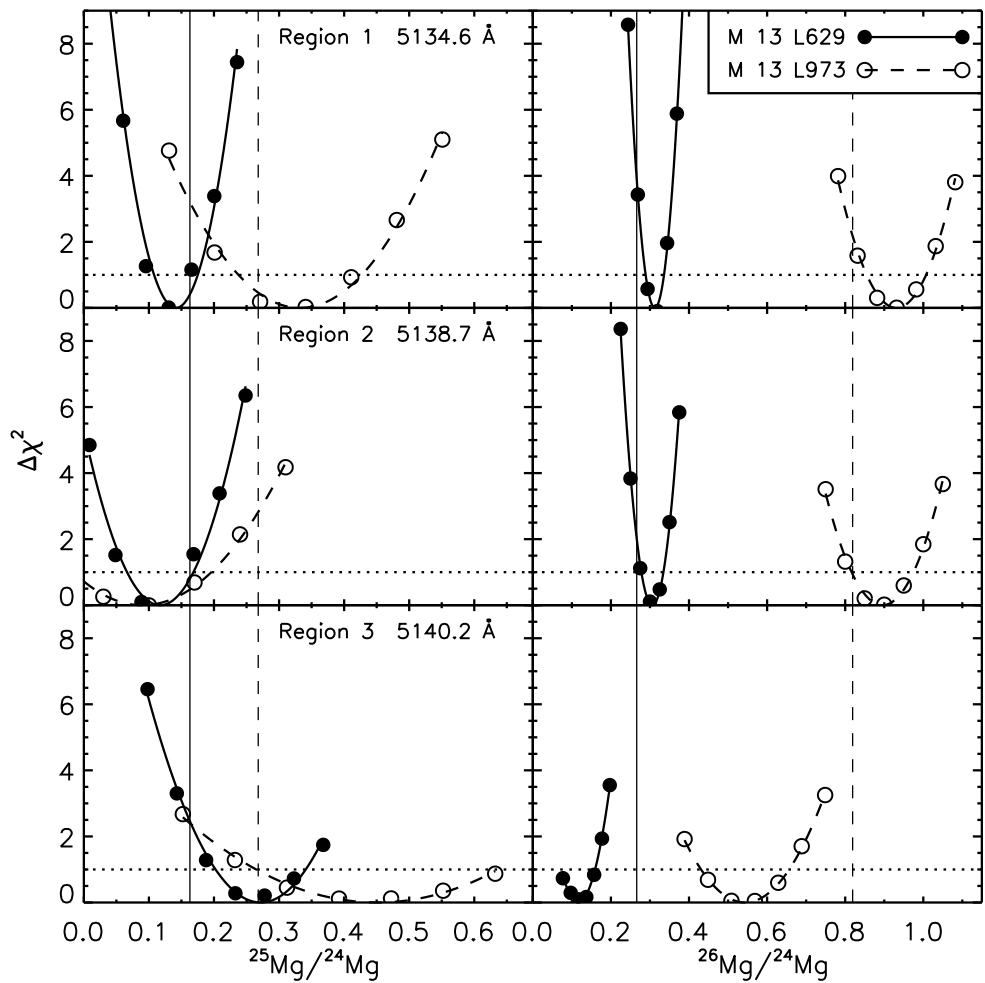}
\caption{Variation in $\Delta\chi^2$ for M 13 L629 and M 13 L973 for 
$^{25}$Mg/$^{24}$Mg ({\it left panels}) and $^{26}$Mg/$^{24}$Mg 
({\it right panels}).
The upper, middle, and lower panels show the $\chi^2$ variation for Region 1 
(5134.6~\AA), Region 2 (5138.7~\AA), and Region 3 (5140.2~\AA) respectively. 
The line indicating the 1$\sigma$ ($\Delta \chi^2 = 1$) errors is shown. 
Vertical lines show the weighted mean values for M 13 L629 (solid) 
and M 13 L973 (dashed). \label{fig:chi}}
\end{figure}

Following \citet{bevington92} and \citet{nissen99,nissen00}, 
we plot $\Delta\chi^2 = \chi^2 - \chi^2_{min}$
for the ratios $^{25}$Mg/$^{24}$Mg and $^{26}$Mg/$^{24}$Mg 
(see Figure \ref{fig:chi}). The $\Delta\chi^2 = 1$ represents the corresponding
1$\sigma$ confidence limit for determining $^{25}$Mg/$^{24}$Mg or 
$^{26}$Mg/$^{24}$Mg. For each recommended line in each star, we paired an 
uncertainty to the optimized ratio $^{25}$Mg/$^{24}$Mg or 
$^{26}$Mg/$^{24}$Mg. The final ratio $^{24}$Mg:$^{25}$Mg:$^{26}$Mg was 
calculated based on the weighted mean. As noted in previous investigations
of Mg isotope ratios, the ratio $^{25}$Mg/$^{24}$Mg 
is less certain than $^{26}$Mg/$^{24}$Mg since 
$^{26}$MgH is less blended with the strong $^{24}$MgH line. Furthermore, 
isotope ratios for Regions 2 and 3 are less accurate than for Region 1. 
Although we calculated the formal statistical errors, these numbers are
small and ignore systematic errors such as continuum fitting, macroturbulence,
and blends. Figures \ref{fig:region1} and \ref{fig:region23} show that the
uncertainties in the ratios are probably at or below the level b $\pm$ 5 or c
$\pm$ 5 when expressing the ratio as 
$^{24}$Mg:$^{25}$Mg:$^{26}$Mg=(100$-$b$-$c):b:c. 
For Region 1 in M 13 L 70, when we lower the continuum level by 1\%, the
best fitting ratio is 49:11:40 (originally 48:14:38). When we change the 
macroturbulence from 7.5 \kms~to 8.0 \kms, the best fitting ratio is 
48:14:38. For Region 2 in M 13 L598, when we lower the continuum 
by 1\% and change the macroturbulence from 6.5 \kms~to 7.0 \kms, 
we measure ratios of 81:6:13 and 84:9:7 respectively (originally
80:10:10).
Note that the derived isotope 
ratios are insensitive to the adopted stellar parameters and appear immune
to non-LTE effects and/or inadequacies in the model atmospheres 
\citep{mghdwarf,mghhyades}. 

In Table \ref{tab:iso} we present the Mg isotope ratios for the program
stars. In this table we also
compare our values with previous determinations. Our superior data, higher
S/N and higher spectral resolution, confirm the pioneering results obtained
by Shetrone. While \citet{shetrone96b} was unable to distinguish between 
$^{25}$Mg and $^{26}$Mg, we note that our ratios $^{25+26}$Mg/$^{24}$Mg are 
in excellent agreement. For the comparison field star, 
HD 141531, the isotope ratio measured from the Subaru data is in excellent 
agreement with the value from the VLT data. 

In the spectra of dwarfs, previous studies of Mg isotope ratios 
(e.g., \citealt{gl2000,mghdwarf,mghhyades}) show that the best fitting 
synthesis to the 5134.6~\AA~line generally predicts a weaker 5134.2~\AA~line
than is observed. 
(As mentioned earlier, \citet{tl80} showed that the 5134.2~\AA~line does not 
provide a reliable isotope ratio.) 
In some cases, the predicted strength of 5134.2~\AA~is very 
close to the observed strength. For cases in 
which the synthetic spectra underpredict the 
observed strength, a weak line not included in the line list could account 
for the small mismatch. However in giants, the mismatch appears to be in the
opposite direction. As shown in Figure \ref{fig:region1}, the synthetic spectra 
overpredict the strength of the MgH 5134.2~\AA~line. Our linelist includes
lines of C$_2$ at 5134.3 and 5134.7~\AA. Our estimate of the C abundance 
was based on the weak C$_2$ lines near 5135.6~\AA~which gave [C/Fe] $\simeq$ 
$-$0.2. Even if we exclude the C$_2$ lines at 5134.3 and 5134.7~\AA, the 
syntheses continue to overpredict the strength of the 5134.2~\AA~line. 

On closer inspection of the NGC 6752 giants, we note that the 
magnitude of the mismatch appears to be a 
function of luminosity with the discrepancy being largest at the tip of the
RGB (\teff~$\simeq$~3900K) and disappearing by \teff~$\simeq$~4400K. Our M 13 and M 71
giants are all located at the tip of the RGB where the discrepancy reaches a maximum. 
We note that a best fit to the 5134.2~\AA~line could be achieved with a reduced 
amount of $^{26}$Mg relative to the 5134.6~\AA~line. However, substantial 
amounts of $^{25}$Mg and $^{26}$Mg would still be required. 

\section{Discussion}
\label{sec:discussion}

\subsection{Elemental abundance ratios in M 13}

First we will focus upon M 13 since our sample for M 71 consists of only 1 
star. Of the elemental abundance ratios that we have measured, [Al/Fe] exhibits 
the largest amplitude. Though we have only observed 4 giants, the [Al/Fe] 
ratios vary by more than an order of magnitude (1.04 dex). This variation 
far exceeds the measurement uncertainty. As a comparison, 
\citet{sneden04} measured Al abundances in a sample of 18 bright giants. 
Excluding the 2 stars for which only upper limits are available, the ratio 
[Al/Fe] varies by roughly 1.2 dex. Therefore, our 4 stars 
span almost the full range of the star-to-star abundance variation for [Al/Fe]. 
Our most Al-rich star, M 13 L973, is also one of the 
most Al-rich stars in the Sneden sample. However, our most
Al-poor star, M 13 L598, is still roughly 0.2 dex more Al-rich than the most 
Al-poor star in the Sneden sample. 
Our ratios of [Na/Fe] vary by almost a factor of 3 (0.47 dex) which again 
greatly exceeds the measurement uncertainty. \citet{sneden04} measured Na 
abundances in 35 bright giants and found the ratio [Na/Fe] varies by 0.95 
dex. Our ratios of [Mg/Fe] vary by a factor of two (0.28 dex) which
is greater than the measurement uncertainty. \citet{sneden04} 
measured [Mg/Fe] in 18 stars and found the ratio varies by 0.57 dex. Comparing
the dispersion in our [Na/Fe] and [Mg/Fe] ratios with the Sneden sample, our
stars span a considerable range of the abundance distribution. However, 
Sneden's M 13 sample contains 
several stars that are more enhanced in Na and Mg as well as several that 
are more depleted in Na and Mg. Even within our small sample, we recover the
Mg-Al anticorrelation and find that Na is correlated with Al in accord with
previous studies of this cluster. 

\subsection{Mg isotope ratios in M 13}

For M 13, we confirm Shetrone's finding that the ratio 
$^{24}$Mg:$^{25}$Mg:$^{26}$Mg varies considerably from star to star. Our
measured ratios $^{25+26}$Mg/$^{24}$Mg are very similar to those 
measured by Shetrone. We find 
that the lowest ratio is 78:11:11 while the highest ratio is 48:13:39. Note
that the lowest observed ratio in NGC 6752 was 84:8:8 and the highest 
observed ratio was 53:9:39. Although the M 13 sample is limited, the extremes
in the isotope ratios appear to have similar values as for NGC 6752. 
\citet{shetrone96b} measured Mg isotope ratios in 6 stars and based on 
these values, we made sure that we observed the stars exhibiting the 
lowest (M 13 L598) and highest (M 13 L70) isotope ratios. 

In M 13 L598, the star with the lowest ratio, we find equal
contributions from $^{25}$Mg and $^{26}$Mg such that $^{25}$Mg = $^{26}$Mg
or equivalently, $^{25}$Mg/$^{24}$Mg = $^{26}$Mg/$^{24}$Mg. 
In NGC 6752, we also found a similar result with 
$^{25}$Mg $\simeq$ $^{26}$Mg in stars with low isotope ratios. For the 
remaining stars in M 13, we do not find $^{25}$Mg = $^{26}$Mg. Instead, the 
contribution from $^{26}$Mg exceeds that from $^{25}$Mg. Such a result would 
again appear to be similar to NGC 6752. Another aspect of the Mg isotope 
ratios common to both M 13 and NGC 6752 is that the contribution of $^{25}$Mg 
to the total Mg abundance is essentially constant from star-to-star with 
$^{25}$Mg/$^{24+25+26}$Mg $\simeq$ 0.13. The isotope ratio 
$^{24}$Mg:$^{25}$Mg:$^{26}$Mg varies due to the changing 
contributions of $^{24}$Mg and $^{26}$Mg, i.e., 
$^{24}$Mg/$^{24+25+26}$Mg and $^{26}$Mg/$^{24+25+26}$Mg differ from
star-to-star in NGC 6752 and M 13. 

In Y03, we noted that at one extreme of the star-to-star abundance variations 
are stars whose compositions [O/Fe], [Na/Fe], 
[Mg/Fe], and [Al/Fe] are essentially 
identical to field stars at the same metallicity, [Fe/H]. 
These are the stars with high O, high Mg, low Na, and low Al 
and we refer to them as ``normal stars''. 
At the other extreme of the abundance variations are the stars 
with high Na, high Al, low O, and low Mg. We referred to these stars as 
``polluted'' in anticipation that proton capture nucleosynthesis can 
produce O-poor, Na-rich, Mg-poor, and Al-rich gas. 
The pollution may have occurred via either the evolutionary or 
primordial scenario. 
Figure 13.8 in \citet{sneden04} shows the O-Na anticorrelation
and the Na-Al correlation. In this Figure, the field stars clearly
occupy one end of the distribution defined by the cluster stars, i.e.,
the region in which the ``normal'' cluster stars are located. 
This Figure, and other large samples of field giants, confirm that 
our single comparison star HD 141531 is representative of field stars. 

Theoretical yields from metal-poor supernovae predict
very small amounts of $^{25}$Mg and $^{26}$Mg relative to $^{24}$Mg, e.g., 
$^{24}$Mg:$^{25}$Mg:$^{26}$Mg $\simeq$ 98:1:1 \citep{woosley95,chieffi04}. 
Based on these yields, Galactic chemical evolution models 
predict $^{24}$Mg:$^{25}$Mg:$^{26}$Mg $\simeq$ 98:1:1 in the range 
$-4 <$ [Fe/H] $< -1$ \citep{timmes95,goswami00,alc01}. 
We stress that in calculating these predictions, the isotopes of Mg are 
assumed to be synthesized solely by Type II supernovae. 
The ``normal'' 
stars in NGC 6752 had ratios of $^{25}$Mg/$^{24}$Mg and $^{26}$Mg/$^{24}$Mg that 
exceeded predictions from metal-poor supernovae as well as the ratios
observed in field stars at the same metallicity (e.g., 
\citealt{gl2000} and \citealt{mghdwarf}). 
Applying the same criteria, M 13 L598 would be a ``normal'' star whereas 
M 13 L629, M 13 L973, and M 13 L70 are ``polluted'' with L70 being the most
``polluted''. Again, we find that the Mg isotope ratios in M 13 parallel those
found in NGC 6752. 
The ``normal'' star M 13 L598 has a Mg isotope ratio
$^{24}$Mg:$^{25}$Mg:$^{26}$Mg = 78:11:11 that exceeds predictions by an 
order of magnitude and these isotope ratios also exceed field stars at 
the same metallicity. At the metallicity of M 13, a 
typical field halo star such as Gmb 1830 (a subdwarf) 
has $^{24}$Mg:$^{25}$Mg:$^{26}$Mg = 94:3:3 \citep{tl80}. 
Our comparison field star, HD 141531, has a ratio 91:4:6. 
In NGC 6752 and M 13, we find that
``normal'' stars have isotope ratios of Mg that differ from field stars
at the same metallicity despite the similarity in elemental abundance
ratios [O/Fe], [Na/Fe], [Mg/Fe], and [Al/Fe]. 

\subsection{Mg isotopic abundances in M 13}

\begin{figure}
\epsscale{0.8}
\plotone{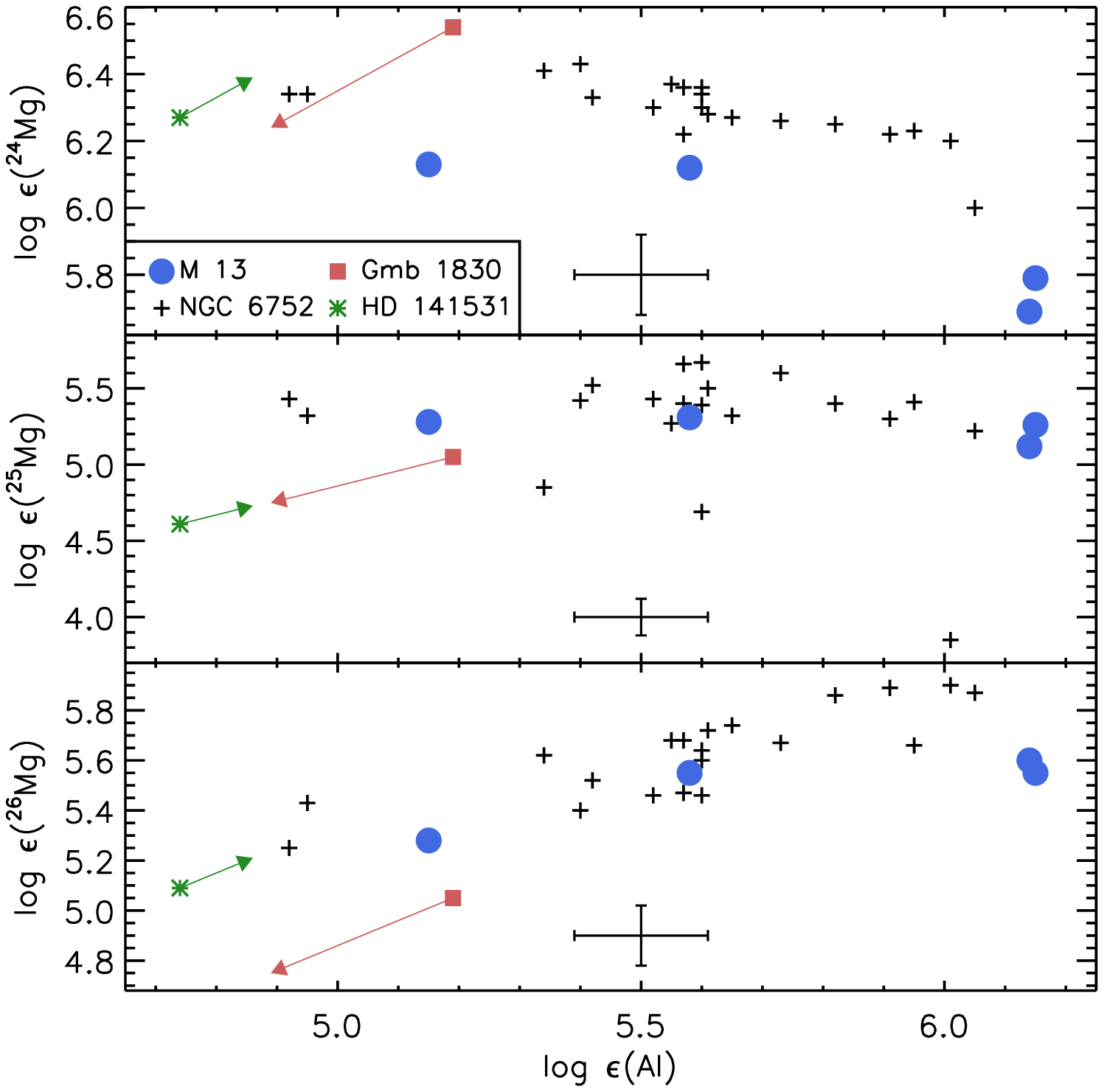}
\caption{The abundances of $^{24}$Mg, $^{25}$Mg, and $^{26}$Mg versus Al.
The filled blue circles represent M 13 while the black plus signs represent
NGC 6752 (Y03). The field stars Gmb 1830 (filled red square: Mg isotope ratio 
from \citealt{tl80} and abundance ratios 
from \citealt{fulbright00}) and HD 141531 (green
asterisk) are overplotted. A representative error bar is shown 
in each panel. The arrows show the field star abundances if we adjust for the
metallicity differences between M 13 and the field stars assuming that 
$\Delta\log~\epsilon$(species) = $\Delta\log~\epsilon$(Fe). \label{fig:almg}}
\end{figure}

In Figure \ref{fig:almg}, we plot the Mg isotopic abundances versus the Al
abundance. The Mg isotopic abundances were derived by combining
the Mg isotope ratios from molecular MgH lines with the Mg elemental 
abundance from Mg\,{\sc i} lines. We note that the $^{24}$Mg 
abundance decreases with increasing Al abundance and that the total
spread in the $^{24}$Mg abundance is a factor of 2.75 which is greater 
than the measurement uncertainty. The $^{25}$Mg abundance appears constant
over the 1.0 dex range in [Al/Fe]. The $^{26}$Mg abundance increases 
with increasing Al abundance. The total dispersion in the $^{26}$Mg abundance 
is a factor of 2.1 which exceeds the measurement uncertainty. 
In Figure \ref{fig:almg} we overplot the results for NGC 6752. 
For M 13 and NGC 6752, an identical behavior is found 
between the Mg isotopic abundances and Al 
abundance. For both clusters, $^{24}$Mg is anticorrelated with Al,
$^{25}$Mg is not correlated with Al, and 
$^{26}$Mg is correlated with Al. 

In this Figure, we also include the
abundances for the comparison field star HD 141531 and the
well studied subdwarf Gmb 1830. (For Gmb 1830 we took the Mg isotope
ratio from \citealt{tl80} and the Mg and Al 
elemental abundances from \citealt{fulbright00}.)
If we adjust the abundances for the small metallicity differences between 
M 13 and the field stars assuming 
$\Delta\log~\epsilon$(species) = $\Delta\log~\epsilon$(Fe), the field stars
have essentially identical Mg isotopic abundances and Al elemental abundances. 
With or without this metallicity correction, the field stars clearly lie at 
one end of the isotopic distribution defined by the cluster stars as 
previously seen in NGC 6752. In Figure \ref{fig:almg}, we can also see that 
the globular clusters have higher isotopic abundances for $^{25}$Mg and
$^{26}$Mg relative to field stars at the same Al abundance. 

\subsection{Elemental abundance ratios and Mg isotope ratios in M 71}

For M 71, our measurements of Mg isotope ratios and elemental abundances 
are restricted to a single star. \citet{sneden94} found evidence
for a variation of O and Na in a sample of 10 stars, though it was not clear
whether there was an O-Na anticorrelation. \citet{ramirez03} observed a larger
sample of stars and found a clear O-Na anticorrelation and evidence for a
variation in the Al abundance for this cluster. 
Our observed star, M 71 A4, was observed by \citet{sneden94} and had 
one of the higher O abundances though its Na abundance was in the middle
of the distribution. Unfortunately, this star was not observed by 
\citet{ramirez03}. Based on the O abundances 
measured by Sneden, this star may be a ``normal'' 
star but the situation is not clear. 

The Mg isotope
ratio in M 71 A4 is $^{24}$Mg:$^{25}$Mg:$^{26}$Mg = 70:13:17. Note that 
$^{26}$Mg/$^{25}$Mg $>$ 1 and such a ratio does not agree perfectly
with the ``normal'' stars in NGC 6752 and M 13 which tend to have $^{25}$Mg = 
$^{26}$Mg. Therefore we tentatively 
suggest that this star is slightly ``polluted'' based on the O and Na elemental
abundances as well as the Mg isotope ratio. For this star, we again find that 
the ratios $^{25}$Mg/$^{24}$Mg and $^{26}$Mg/$^{24}$Mg exceed those observed
in field stars of comparable metallicity. Finally, we note that the 
contribution of $^{25}$Mg to the total Mg abundance is similar to NGC 6752
and M 13 with $^{25}$Mg/$^{24+25+26}$Mg = 0.13. 

If we assume that the ratio $^{25}$Mg/$^{24+25+26}$Mg 
in M 71 A4 is representative of all stars in this cluster (a constant
ratio for all stars was 
found in the larger samples in M 13 and NGC 6572), 
then such a ratio may be regarded as unusual. The observed and predicted ratio 
$^{25}$Mg/$^{24+25+26}$Mg increases with increasing metallicity 
\citep{timmes95,gl2000,goswami00,alc01,mghdwarf}. 
Even if this ratio in globular clusters exceeds field stars at the same
metallicity, we may expect that in globular clusters the ratio would also
increase with increasing metallicity. Instead, it seems that 
the ratio $^{25}$Mg/$^{24+25+26}$Mg does not differ between M 13, NGC 6752, 
and M 71 despite the iron abundance changing from [Fe/H] = $-$1.6 to 
[Fe/H] = $-$0.9 as well as the large dispersions in O-Al abundances. 
Clearly the analyses of additional stars within this cluster (and clusters
with different metallicities) are necessary 
to determine whether there is a spread in the isotope ratio, 
$^{24}$Mg is anticorrelated with Al and/or O, $^{25}$Mg is not correlated 
with O, $^{26}$Mg is correlated with O, and if the lowest isotope ratios
exceed field stars at the same metallicity. 

\subsection{Implications for globular cluster chemical evolution}

\citet{shetrone96b} provided the first measurements of Mg isotope ratios
in a globular cluster. His results for M 13 showed that the ratio 
$^{25+26}$Mg/$^{24}$Mg varied from star-to-star. Of 
great interest was his discovery that the abundant $^{24}$Mg decreases
with increasing Al abundance. Very high temperatures are necessary for
proton-capture on $^{24}$Mg. Shetrone proposed a deep-mixing sequence 
in which Li begins to be destroyed at 10 $\times~10^6$ K. At 20 
$\times~10^6$ K, C begins to decrease, N increases, and the ratio 
$^{12}$C/$^{13}$C decreases. At 30 $\times~10^6$ K, O begins to decrease,
Na begins to increase, and C and N continue fall and rise respectively. 
The Al abundance increases slightly due to destruction of $^{25}$Mg and
$^{26}$Mg at 40 $\times~10^6$ K. Finally, Al increases significantly as 
$^{24}$Mg decreases at 70 $\times~10^6$ K. Based on Shetrone's isotope ratios 
(and measurements of other elemental abundance ratios), 
various schemes were explored in which the
abundance variations could be produced through the evolutionary scenario,
the primordial scenario, and combinations of the two (e.g., 
\citealt{langer97}, \citealt{denissenkov97,denissenkov98}, and 
\citealt{weiss00}). 

Observations by \citet{gratton01} showed that 
the anticorrelations of O-Na and Mg-Al exist even
in main sequence stars and early subgiants in NGC 6752. Subsequent measurements
have shown a similar result for M 71 \citep{ramirez03} and 
M 13 \citep{cohen05}. The range
and amplitude of the abundance variations of O-Al do not vary from the
main sequence to the tip of the red giant branch within a given 
cluster. This behavior 
differs from the measurements of C and N abundances that vary with 
evolutionary status \citep{ss91}. 
Al variations in unevolved stars demonstrate that the abundance 
variations cannot be due primarily to internal mixing and nucleosynthesis 
because main sequence stars and early subgiants have internal temperatures 
too low to run the Ne-Na or Mg-Al chains. Therefore, the present cluster 
members must have formed
from inhomogeneous gas or accreted such material. 

Measurements of Mg isotope ratios in NGC 6752 led Y03 to suggest a revised
primordial scenario involving two generations of IM-AGBs. 
``Normal stars'' showed Mg isotope ratios that exceeded both the 
predictions from metal-poor supernovae as well as the ratios observed 
in field stars at the same metallicity. Y03 suggested that a prior
generation of metal-poor IM-AGBs can raise the low amounts of $^{25}$Mg and 
$^{26}$Mg provided by metal-poor supernovae to the much higher levels observed 
in the ``normal'' stars. In order to preserve the highly 
uniform abundances of Fe, Ca, Ni, etc \citep{67522}, 
the ejecta from these metal-poor IM-AGBs must be thoroughly mixed with
the ejecta from metal-poor supernovae prior to the formation of the present
cluster members. As seen in Shetrone's M 13 sample, 
the large Al enhancement was accompanied by a decrease in $^{24}$Mg 
for the NGC 6752 giants. Destruction of $^{24}$Mg only takes place at very 
high temperatures such as in AGB stars of the highest mass. Y03 suggested 
that IM-AGBs pollute the environment from which the present generation of
stars formed, an idea first proposed by \citet{cottrell81}. 
In IM-AGBs, hydrogen-burning at the base of the
convective envelope, so-called hot bottom burning (HBB), may produce the 
observed C to Al abundance patterns. 
Y03 also showed that the Fe abundance was identical in all
stars within the measurement error. So the source of the pollutants must have 
the same Fe abundance as the present generation of stars. The O abundance 
varied by nearly an order of magnitude within NGC 6752. In the event that the
pollutants contain no O, then the factor of 10 decrease in the O abundance 
of the most ``polluted'' stars relative to the ``normal'' stars is only possible
via a mix of 90\% pollutants to 10\% normal material. In addition to the
dominant primordial component, an evolutionary component is essential to 
account for the C \citep{ss91} and Li \citep{grundahl02} destruction with 
increasing luminosity.

In M 13, we find that stars with elemental abundance ratios in agreement 
with field stars at the same metallicity, i.e., ``normal'' stars, have Mg 
isotope ratios that exceed those seen in field stars. A similar result was 
seen in NGC 6752. 
Therefore, we reiterate
the suggestion from Y03 that 
a prior generation of metal-poor IM-AGBs
are needed to produce the high $^{25}$Mg/$^{24}$Mg and $^{26}$Mg/$^{24}$Mg 
seen in ``normal'' stars of M 13 and NGC 6752. 

The principal result of this study is that the neutron-rich Mg isotopes do not 
necessarily have the 
same abundance in M 13, i.e., $^{25}$Mg/$^{24}$Mg $\neq$ $^{26}$Mg/$^{24}$Mg. 
Instead, the isotopic abundance of $^{25}$Mg is constant
across the 1.0 dex range in Al while the isotopic abundance of $^{26}$Mg 
is correlated with Al. We also confirm Shetrone's 
result that $^{24}$Mg decreases with increasing
Al abundance. 
These results for M 13 are identical to the behavior of the
Mg isotopic abundances in NGC 6752 suggesting that the same mechanism is 
responsible for the isotopic and elemental abundance variations seen
in both clusters. That is, two generations of IM-AGB stars pollute the
cluster as just described. 

The similarity in the correlations between Mg isotopes and Al abundances in 
M 13 and NGC 6752 may be regarded as surprising. 
If the Al variations arise from the IM-AGBs, then it may be reasonable to 
expect a large cluster-to-cluster variation of the pollution from such 
stars. Therefore an additional aspect to the IM-AGB scenario may be
required. Perhaps the IM-AGBs were of a very similar mass in both clusters. 

For M 71, our one star has a Mg isotope ratio 
$^{24}$Mg:$^{25}$Mg:$^{26}$Mg = 70:13:17, i.e., 
$^{25}$Mg/$^{24}$Mg $\neq$ $^{26}$Mg/$^{24}$Mg. We also note that
$^{26}$Mg/$^{25}$Mg $>$ 1, a ratio that is always seen in the 
``polluted'' stars of M 13 and NGC 6752. For this star, we find 
$^{25}$Mg/$^{24+25+26}$Mg = 0.13 and such a ratio was also 
seen in every star in M 13 and NGC 6752. M 71 A4 has ratios 
$^{25}$Mg/$^{24}$Mg and $^{26}$Mg/$^{24}$Mg that exceed field stars at
the same metallicity. Such a difference may again be interpreted as resulting
from the pollution of the proto-cluster gas by a prior generation
of metal-poor IM-AGBs. Clearly additional measurements of Mg isotope 
ratios in O-rich and O-poor stars in M 71 are required to see if the
Mg isotope ratio varies from star to star, the Mg isotopic abundances are
correlated with O, and if the contribution of $^{25}$Mg to the total Mg
abundance is constant. We speculate that the highest ratios of 
$^{26}$Mg/$^{24}$Mg in M~71 will not be as extreme as those seen in M 13 and 
NGC 6752 since the amplitude of the Al variation in M 71 is smaller. 

If differing degrees of pollution from IM-AGBs of the same generation as the 
present cluster members is the correct mechanism for producing the O-Al 
abundance variations, then 
Fluorine may be expected to exhibit a star-to-star abundance variation. 
During HBB in IM-AGBs, destruction of F is predicted in AGB models of the
highest mass \citep{lattanzio04}. 
Recently, \citet{smith05} showed that F varied from star-to-star in M 4 and 
that the F abundance was correlated with O and anticorrelated with Na and Al. 
Such variations of F lend further support to the AGB pollution scenario. 

While the AGB pollution scenario offers a plausible qualitative explanation for
the light element abundance anomalies, 
quantitative tests reveal several problems. The observed ratios of 
$^{25}$Mg/$^{24}$Mg and $^{26}$Mg/$^{24}$Mg in stars with O depletions 
are significantly lower than predictions from AGB models 
\citep{denissenkov03}. The low C and O abundances and high N cannot be
produced in AGB models \citep{denissenkov04}. In the detailed
globular cluster chemical evolution model by \citet{fenner04}, 
O was not depleted, Mg was produced rather than destroyed, C+N+O was
not constant, and $^{25}$Mg should be correlated with $^{26}$Mg. 
These inconsistencies between the observed and predicted compositions
highlight that our understanding of stellar nucleosynthesis and globular
cluster chemical evolution are incomplete. 
While the details are model dependent, \citet{busso01} suggest that metal-poor 
IM-AGBs may run the $s$-process. 
\citet{67522} found evidence for 
slight abundance variations for Y, Zr, and Ba with each element showing a 
statistically significant but small correlation with 
Al. Therefore, the stars responsible for the synthesis of the Al anomalies 
may have also run the $s$-process.  
Indeed, 
\citet{ventura05} caution that the choice of mass-loss rates and treatment 
of convection greatly affect the calculated 
AGB yields to such an extent that the
predictive power of the AGB models is limited.

New constraints on the source of the abundance anomalies have come from 
observations of Li in main sequence stars in NGC 6752. 
\citet{pasquini05} have shown that Li is correlated with O
and anticorrelated with Na in these unevolved stars. 
Such a behavior is expected given that the
Ne-Na cycle operates at temperatures far exceeding those required 
to destroy Li. \citet{pasquini05} emphasize that Li is observed in even
the most ``polluted'' stars with log $\epsilon$(Li) $\simeq$ 2.0, an amount
only 2-3 times lower than in stars with the highest Li abundance. Since
O and Na vary by roughly an order of magnitude, this implies that Li must also 
have been synthesized by the stars responsible for the abundance variations. 
While \citet{pasquini05} also recognize that the yields from IM-AGBs cannot
reproduce all the observed abundances, they note that the \citet{ventura02} 
Li yields for metal-poor IM-AGBs are in good agreement with the observations. 
\citet{pasquini05} also note that the IM-AGB scenario requires a very large
number of such stars and therefore an unusual initial mass function. 
Furthermore, a large number of white dwarfs could still remain in the cluster 
depending upon the dynamical history of the cluster. 

\section{Concluding remarks}
\label{sec:summary}

Mg isotope ratios are measured in four bright red giants of the globular 
cluster M 13 as well as one bright red giant in the globular cluster M 71. 
We confirm Shetrone's findings that the ratio $^{25+26}$Mg/$^{24}$Mg varies 
from star to star, that $^{25+26}$Mg/$^{24}$Mg $\simeq$ 1 in stars with the
highest Al abundance, and that $^{24}$Mg decreases with increasing Al. 
Since our data have superior resolution and S/N, we 
are able to distinguish the contributions of all three isotopes.
In ``normal'' stars in both NGC 6752 and M 13, the Mg isotope 
ratios $^{25}$Mg/$^{24}$Mg and $^{26}$Mg/$^{24}$Mg 
exceed those observed in field stars at the
same metallicity.
The principal findings are that $^{25}$Mg $\neq$ $^{26}$Mg, that $^{25}$Mg 
is not correlated with Al, and that $^{26}$Mg is correlated 
with Al. 
The behavior of the Mg isotope ratios 
in M 13 is almost identical to that seen 
in NGC 6752 
and suggests that the abundance variations in
both clusters are a result of the same mechanism. 
While yields from AGB models cannot explain all the observed abundances, 
we continue to propose that two generations of IM-AGB stars produce
the observed abundances. A previous generation of metal-poor IM-AGBs
polluted the cluster from which the present generation of stars formed. 
These IM-AGBs are required to produce the high ratios of 
$^{25}$Mg/$^{24}$Mg and $^{26}$Mg/$^{24}$Mg seen in ``normal'' stars 
since metal-poor supernovae do not produce $^{25}$Mg and $^{26}$Mg in 
significant quantities. A second generation of IM-AGBs with the same
metallicity as the cluster are born and evolve. Differing degrees of
pollution from these IM-AGBs produce the observed O-Al variations. 
In M 71, our one star has a contribution of 
$^{25}$Mg identical to that seen in M 13 and NGC 6752, i.e., 
$^{25}$Mg/$^{24+25+26}$Mg = 0.13. Our M 71 star also has $^{26}$Mg $>$ 
$^{25}$Mg, a result only seen in the ``polluted'' stars in M 13 and 
NGC 6752. The isotope ratios $^{25}$Mg/$^{24}$Mg and $^{26}$Mg/$^{24}$Mg
in M 71 also exceed field stars at the same metallicity. 
Finally, all stars in all clusters appear to have the same 
contribution of $^{25}$Mg to the total Mg abundance, i.e., 
$^{25}$Mg/$^{24+25+26}$Mg $\simeq$ 0.13 despite the considerable 
differences in [Fe/H], [O/Fe], and [Al/Fe]. 
Having shown that the Mg isotope ratios are similar in clusters
with large O-Al variations, additional measurements in clusters that
exhibit mild O-Al variations are required to further our understanding
of the globular cluster star-to-star abundance variations. 

\acknowledgments

This research has made use of the SIMBAD database,
operated at CDS, Strasbourg, France and
NASA's Astrophysics Data System. We thank the anonymous referee 
for helpful comments. 
DY thanks Bruce Carney for a thorough review of a draft of this paper.
DLL acknowledges support from the Robert A.\ Welch Foundation of Houston, Texas.
This research was 
supported in part by NASA through the American Astronomical Society's Small 
Research Grant Program.

\clearpage

\clearpage

\clearpage

\clearpage

\clearpage

\input{tab1}
\input{tab6}
\input{tab2}
\input{tab3}
\input{tab4}
\input{tab5}

\end{document}

%% file: tab1.tex
\begin{deluxetable}{lcccccccc} 
\tabletypesize{\footnotesize}
\tablecolumns{9} 
\tablewidth{0pc} 
\tablecaption{Exposure times and stellar parameters.\label{tab:param}}
\tablehead{ 
\colhead{Star} &
\colhead{Alternative} &
\colhead{Exposure} &
\colhead{S/N\tablenotemark{a}} &
\colhead{\teff} &
\colhead{log $g$} &
\colhead{$\xi_t$} &
\colhead{Macro} &
\colhead{[Fe/H]} \\
\colhead{} &
\colhead{Name} &
\colhead{Time (min)} &
\colhead{5140\AA} &
\colhead{K} &
\colhead{} &
\colhead{km s$^{-1}$} &
\colhead{km s$^{-1}$} &
\colhead{}
}
\startdata
M 13 L598 & & 60 & 141 & 3900 & 0.0 & 2.25 & 6.25 & $-$1.56 \\
M 13 L629 & & 50 & 129 & 3950 & 0.2 & 2.25 & 7.50 & $-$1.63 \\
M 13 L70 & II-67 & 60 & 130 & 3950 & 0.3 & 2.25 & 7.50 & $-$1.59 \\
M 13 L973 & I-48 & 64 & 126 & 3920 & 0.3 & 2.35 & 7.00 & $-$1.61 \\
M 71 A4 & & 50 & 140 & 4100 & 0.8 & 2.05 & 4.50 & $-$0.91 \\
HD 141531 & & 10 & 217 & 4273 & 0.8 & 1.90 & 5.00 & $-$1.72 \\
\enddata

\tablenotetext{a}{S/N values are per pixel (4 pixels per resolution element).}

\end{deluxetable}

%% file: tab6.tex
\begin{deluxetable}{lcccccccc} 
\tabletypesize{\scriptsize}
\tablecolumns{9} 
\tablewidth{0pc} 
\tablecaption{Equivalent widths for program stars [Online Table].\label{tab:ew}}
\tablehead{ 
\colhead{\AA} &
\colhead{Species} &
\colhead{$\chi$ (eV)} &
\colhead{log $gf$} &
\colhead{M 13 L598} &
\colhead{M 13 L629} &
\colhead{M 13 L70} &
\colhead{M 13 L973} &
\colhead{M 71 A4}
}
\startdata
6300.30 & 8.0 & 0.00 & $-$9.75 & 63.7 & 44.3 & \ldots & \ldots & 86.0 \\
6363.78 & 8.0 & 0.02 & $-$10.25 & 30.6 & 17.6 & \ldots & \ldots & 48.0 \\
4982.83 & 11.0 & 2.10 & $-$0.91 & 54.3 & 71.0 & 80.3 & 66.3 & 81.6 \\
5682.65 & 11.0 & 2.10 & $-$0.71 & 75.2 & 90.8 & 104.8 & 100.8 & 139.2 \\
5688.22 & 11.0 & 2.10 & $-$0.40 & 100.1 & 121.0 & 134.0 & 126.3 & 150.6 \\
6154.23 & 11.0 & 2.10 & $-$1.56 & 20.3 & 28.8 & 38.4 & 34.1 & 66.1 \\
6160.75 & 11.0 & 2.10 & $-$1.26 & 33.3 & 45.0 & 61.3 & 53.4 & 93.0 \\
5711.09 & 12.0 & 4.35 & $-$1.73 & 101.6 & 103.5 & 79.2 & 78.6 & 138.5 \\
6318.71 & 12.0 & 5.11 & $-$1.97 & 17.5 & 17.1 & 13.8 & 11.1 & 61.4 \\
6319.24 & 12.0 & 5.11 & $-$2.20 & 11.3 & \ldots & \ldots & \ldots & 50.4 \\
6696.02 & 13.0 & 3.14 & $-$1.57 & 24.8 & 51.2 & 100.5 & 96.7 & 93.0 \\
6698.67 & 13.0 & 3.14 & $-$1.89 & 13.7 & 27.6 & 63.3 & 64.9 & 60.7 \\
4439.88 & 26.0 & 2.28 & $-$3.00 & 91.3 & 84.7 & \ldots & \ldots & \ldots \\
4802.88 & 26.0 & 3.69 & $-$1.53 & 67.7 & 62.7 & 65.4 & 63.0 & 80.5 \\
4817.78 & 26.0 & 2.22 & $-$3.53 & 82.6 & 73.9 & 81.2 & 80.8 & 111.5 \\
4839.55 & 26.0 & 3.27 & $-$1.84 & 93.8 & 86.5 & 87.5 & \ldots & \ldots \\
4896.44 & 26.0 & 3.88 & $-$1.90 & 29.9 & 26.8 & 24.9 & 28.2 & 60.7 \\
4930.32 & 26.0 & 3.96 & $-$1.20 & 67.5 & 61.1 & 62.7 & 57.8 & 114.1 \\
4962.57 & 26.0 & 4.18 & $-$1.20 & 44.2 & 42.7 & 45.3 & \ldots & 68.7 \\
4969.92 & 26.0 & 4.22 & $-$0.75 & 73.4 & 67.1 & 69.0 & 64.5 & 119.2 \\
5002.79 & 26.0 & 3.40 & $-$1.44 & 102.4 & 100.1 & 94.1 & 93.2 & \ldots \\
5029.62 & 26.0 & 3.42 & $-$1.90 & 70.7 & 61.2 & 58.6 & 60.7 & \ldots \\
5044.21 & 26.0 & 2.85 & $-$2.03 & \ldots & 113.8 & 112.3 & 107.0 & \ldots \\
5054.64 & 26.0 & 3.64 & $-$1.94 & 48.4 & 39.3 & 40.8 & 38.2 & \ldots \\
5090.77 & 26.0 & 4.26 & $-$0.36 & 90.8 & 83.0 & 83.7 & 78.7 & \ldots \\
5121.64 & 26.0 & 4.28 & $-$0.72 & 67.0 & 59.2 & 62.4 & 56.0 & \ldots \\
5143.72 & 26.0 & 2.20 & $-$3.79 & 91.9 & \ldots & 67.7 & \ldots & \ldots \\
5222.39 & 26.0 & 2.28 & $-$3.68 & 56.4 & 48.4 & 50.9 & \ldots & 95.0 \\
5223.19 & 26.0 & 3.63 & $-$1.80 & \ldots & \ldots & \ldots & \ldots & 70.0 \\
5242.49 & 26.0 & 3.63 & $-$0.98 & 107.4 & 102.1 & 101.1 & 99.0 & 113.1 \\
5253.46 & 26.0 & 3.28 & $-$1.63 & 98.9 & \ldots & \ldots & 92.0 & \ldots \\
5267.28 & 26.0 & 4.37 & $-$1.66 & 16.5 & 14.0 & 14.4 & 11.9 & \ldots \\
5288.53 & 26.0 & 3.69 & $-$1.51 & 72.3 & 67.2 & 65.1 & 64.8 & 82.3 \\
5298.78 & 26.0 & 3.64 & $-$2.03 & \ldots & \ldots & \ldots & \ldots & 66.7 \\
5321.11 & 26.0 & 4.43 & $-$1.11 & 31.2 & 27.2 & 31.1 & 31.7 & 60.2 \\
5326.14 & 26.0 & 3.57 & $-$2.13 & 37.4 & 31.9 & 37.2 & 35.2 & 66.2 \\
5466.40 & 26.0 & 4.37 & $-$0.57 & 72.4 & 68.0 & 68.3 & 66.6 & 110.4 \\
5487.75 & 26.0 & 4.32 & $-$0.65 & 88.9 & \ldots & \ldots & 85.6 & \ldots \\
5491.84 & 26.0 & 4.18 & $-$2.25 & \ldots & \ldots & \ldots & \ldots & 19.2 \\
5522.45 & 26.0 & 4.21 & $-$1.40 & 34.0 & 28.3 & 36.2 & 31.3 & 56.7 \\
5525.54 & 26.0 & 4.23 & $-$1.15 & 43.6 & 40.3 & 44.4 & \ldots & 65.6 \\
5554.90 & 26.0 & 4.55 & $-$0.38 & 69.7 & 64.5 & 67.1 & 60.6 & 102.2 \\
5560.21 & 26.0 & 4.43 & $-$1.04 & 36.1 & 34.0 & 35.4 & 34.2 & 59.2 \\
5567.39 & 26.0 & 2.61 & $-$2.80 & 109.0 & 102.9 & 104.9 & 98.9 & \ldots \\
5584.77 & 26.0 & 3.57 & $-$2.17 & 44.2 & \ldots & 40.8 & 39.6 & \ldots \\
5618.63 & 26.0 & 4.21 & $-$1.26 & 40.6 & 36.5 & 38.1 & 34.1 & 62.1 \\
5624.02 & 26.0 & 4.39 & $-$1.33 & 34.0 & 28.7 & 34.5 & 33.1 & 60.9 \\
5633.95 & 26.0 & 4.99 & $-$0.12 & 40.5 & 33.7 & 35.7 & 36.8 & 65.9 \\
5635.82 & 26.0 & 4.26 & $-$1.74 & 22.8 & 18.0 & 21.4 & 18.6 & 53.2 \\
5638.26 & 26.0 & 4.22 & $-$0.72 & 73.9 & 68.3 & 72.0 & 67.8 & 91.0 \\
5661.35 & 26.0 & 4.28 & $-$1.82 & 15.4 & 11.8 & 16.2 & 12.0 & 41.3 \\
5679.02 & 26.0 & 4.65 & $-$0.77 & 37.6 & 33.1 & 35.6 & 34.5 & 60.6 \\
5705.47 & 26.0 & 4.30 & $-$1.36 & 29.2 & 22.0 & 26.2 & 25.2 & 50.4 \\
5731.76 & 26.0 & 4.26 & $-$1.15 & 51.9 & 47.4 & 49.4 & 47.6 & 72.0 \\
5741.85 & 26.0 & 4.25 & $-$1.69 & \ldots & \ldots & 20.1 & 20.9 & 44.7 \\
5753.12 & 26.0 & 4.26 & $-$0.71 & 73.0 & 66.4 & 67.9 & 65.6 & \ldots \\
5775.08 & 26.0 & 4.22 & $-$1.31 & 53.4 & 48.7 & 49.9 & 44.9 & 72.3 \\
5778.45 & 26.0 & 2.59 & $-$3.48 & 54.8 & 49.6 & 51.5 & 50.8 & 74.7 \\
5784.66 & 26.0 & 3.39 & $-$2.58 & \ldots & \ldots & 36.1 & \ldots & \ldots \\
5816.37 & 26.0 & 4.55 & $-$0.62 & 56.8 & 45.5 & 54.7 & 50.9 & 84.8 \\
5855.09 & 26.0 & 4.60 & $-$1.55 & 11.6 & 8.2 & 10.0 & \ldots & 31.9 \\
5909.97 & 26.0 & 3.21 & $-$2.64 & \ldots & 63.7 & 61.6 & 63.4 & 95.4 \\
5916.25 & 26.0 & 2.45 & $-$2.99 & 107.6 & 100.2 & 103.4 & 102.9 & \ldots \\
6012.21 & 26.0 & 2.22 & $-$4.07 & 59.9 & 48.9 & 53.0 & 50.0 & 75.4 \\
6027.05 & 26.0 & 4.07 & $-$1.11 & 63.1 & 62.1 & 64.2 & 62.3 & 82.8 \\
6082.71 & 26.0 & 2.22 & $-$3.57 & 96.4 & 83.3 & 87.6 & 84.2 & 108.6 \\
6120.24 & 26.0 & 0.91 & $-$5.97 & 64.2 & 50.8 & 55.0 & 52.9 & 81.3 \\
6151.62 & 26.0 & 2.17 & $-$3.30 & \ldots & 106.0 & 108.4 & 111.6 & \ldots \\
6165.36 & 26.0 & 4.14 & $-$1.49 & 36.5 & 33.1 & 37.3 & \ldots & 60.2 \\
6180.20 & 26.0 & 2.73 & $-$2.64 & 101.8 & 96.0 & 92.6 & 94.4 & 106.0 \\
6229.23 & 26.0 & 2.84 & $-$2.85 & 70.8 & 65.4 & 66.4 & 67.2 & 82.3 \\
6232.64 & 26.0 & 3.65 & $-$1.28 & 97.2 & 93.6 & 98.6 & 94.7 & 102.8 \\
6270.22 & 26.0 & 2.86 & $-$2.51 & 94.6 & 87.1 & 89.5 & 84.8 & 103.5 \\
6271.28 & 26.0 & 3.33 & $-$2.76 & 34.7 & 34.3 & 31.5 & 31.6 & 58.9 \\
6301.50 & 26.0 & 3.65 & $-$0.77 & \ldots & \ldots & \ldots & 119.8 & \ldots \\
6336.82 & 26.0 & 3.68 & $-$0.92 & \ldots & \ldots & \ldots & 113.1 & \ldots \\
6344.15 & 26.0 & 2.43 & $-$2.92 & \ldots & 119.2 & 118.0 & 117.8 & \ldots \\
6353.84 & 26.0 & 0.91 & $-$6.48 & 30.6 & 24.5 & 30.4 & 25.3 & 49.2 \\
6355.03 & 26.0 & 2.84 & $-$2.40 & \ldots & 116.1 & \ldots & \ldots & \ldots \\
6408.02 & 26.0 & 3.68 & $-$1.07 & 113.4 & 110.2 & 109.5 & 106.3 & 117.8 \\
6518.36 & 26.0 & 2.83 & $-$2.50 & 98.0 & 90.9 & 89.6 & 88.1 & 104.2 \\
6575.02 & 26.0 & 2.59 & $-$2.73 & \ldots & 115.8 & 112.2 & 113.4 & \ldots \\
6581.21 & 26.0 & 1.48 & $-$4.71 & \ldots & 81.6 & 87.2 & \ldots & 103.2 \\
6609.11 & 26.0 & 2.56 & $-$2.69 & \ldots & 117.6 & 115.8 & 116.0 & \ldots \\
6625.02 & 26.0 & 1.01 & $-$5.37 & \ldots & 107.7 & 114.1 & 112.4 & \ldots \\
6648.08 & 26.0 & 1.01 & $-$5.92 & 70.2 & 59.0 & 64.2 & 67.2 & 81.4 \\
4491.40 & 26.1 & 2.85 & $-$2.68 & \ldots & 78.5 & \ldots & \ldots & \ldots \\
4508.29 & 26.1 & 2.85 & $-$2.31 & \ldots & 102.8 & \ldots & \ldots & \ldots \\
4576.34 & 26.1 & 2.84 & $-$2.82 & \ldots & 80.3 & \ldots & \ldots & \ldots \\
4620.52 & 26.1 & 2.83 & $-$3.08 & 49.5 & 46.1 & 45.0 & 46.1 & \ldots \\
4993.35 & 26.1 & 2.81 & $-$3.67 & 42.7 & 38.5 & 34.9 & 34.4 & 68.5 \\
5100.66 & 26.1 & 2.81 & $-$4.14 & 28.8 & 22.9 & \ldots & \ldots & \ldots \\
5132.67 & 26.1 & 2.81 & $-$3.90 & \ldots & \ldots & \ldots & 15.5 & \ldots \\
5234.62 & 26.1 & 3.22 & $-$2.24 & 86.4 & 80.3 & \ldots & \ldots & 68.8 \\
5264.81 & 26.1 & 3.23 & $-$3.19 & 32.7 & 30.9 & 29.2 & 36.4 & 30.6 \\
5284.10 & 26.1 & 2.89 & $-$3.01 & 58.5 & \ldots & \ldots & \ldots & \ldots \\
5534.83 & 26.1 & 3.25 & $-$2.77 & 70.2 & 65.8 & \ldots & 58.6 & \ldots \\
5991.38 & 26.1 & 3.15 & $-$3.56 & 23.0 & 20.9 & 25.5 & 24.7 & 28.8 \\
6084.11 & 26.1 & 3.20 & $-$3.81 & 22.5 & 12.4 & 15.8 & 17.6 & 15.8 \\
6149.26 & 26.1 & 3.89 & $-$2.72 & 23.4 & 21.2 & 18.8 & 23.5 & \ldots \\
6247.56 & 26.1 & 3.89 & $-$2.33 & 30.2 & 29.9 & 26.5 & 31.4 & \ldots \\
6369.46 & 26.1 & 2.89 & $-$4.25 & 15.2 & 13.6 & 14.5 & 16.3 & 24.1 \\
6416.92 & 26.1 & 3.89 & $-$2.74 & 23.7 & 22.7 & 18.6 & 22.4 & 32.1 \\
6432.68 & 26.1 & 2.89 & $-$3.71 & 36.1 & 32.5 & 30.4 & 30.9 & 32.6 \\
6456.38 & 26.1 & 3.90 & $-$2.08 & 46.5 & 44.6 & 44.5 & 44.4 & 41.2 \\
6516.08 & 26.1 & 2.89 & $-$3.45 & 48.5 & 48.5 & 44.4 & 40.5 & 40.1 \\

\enddata

\end{deluxetable}

%% file: tab2.tex
\begin{deluxetable}{lcccccc} 
\tabletypesize{\footnotesize}
\tablecolumns{8} 
\tablewidth{0pc} 
\tablecaption{Stellar parameters comparison.\label{tab:comp}}
\tablehead{ 
\colhead{Star} &
\colhead{Alternative} &
\colhead{\teff} &
\colhead{log $g$} &
\colhead{$\xi_t$} &
\colhead{[Fe/H]} & 
\colhead{Source}\\
\colhead{} &
\colhead{Name} &
\colhead{K} &
\colhead{} &
\colhead{km s$^{-1}$} &
\colhead{} &
\colhead{}
}
\startdata
M 13 L598 & & 3925 & 0.00 & 2.36 & $-$1.56 & 1 \\
          & & 3900 & 0.00 & 2.25 & $-$1.48 & 2 \\
          & & 3900 & 0.30 & \ldots & $-$1.55 & 3 \\
M 13 L629 & & 3975 & 0.10 & 2.36 & $-$1.63 & 1\\
          & & 3950 & 0.20 & 2.25 & $-$1.46 & 2 \\
          & & 4010 & 0.36 & \ldots & $-$1.62 & 3 \\
M 13 L70 & II-67 & 3950 & 0.10 & 2.14 & $-$1.59 & 1 \\
         &       & 3950 & 0.30 & 2.25 & $-$1.51 & 2 \\
         &       & 3900 & 0.37 & \ldots & $-$1.49 & 3 \\
         &       & 3900 & 0.45 & 1.90 & $-$1.30\tablenotemark{a} & 4 \\
M 13 L973 & I-48 & 3925 & 0.10 & 2.14 & $-$1.61 & 1 \\
          &      & 3920 & 0.30 & 2.35 & $-$1.45 & 2 \\
          &      & 3950 & 0.34 & \ldots & $-$1.57 & 3 \\
M 71 A4 & & 4100 & 0.80 & 2.05 & $-$0.91 & 1 \\
        & & 4100 & 0.80 & 2.25 & $-$0.76 & 2 \\
        & & 4100 & 0.80 & 2.25 & $-$0.76 & 5 \\
\enddata

\tablenotetext{a}{There is a 0.3 dex discrepancy between [Fe/H]$_{\rm I}$ and [Fe/H]$_{\rm II}$ with [Fe/H]$_{\rm I}$ = $-$1.45}

\tablerefs{
1 = This study;
2 = \citet{shetrone96a,shetrone96b};
3 = \citet{sneden04a};
4 = \citet{cohen05}; 
5 = \citet{sneden94}
}

\end{deluxetable}

%% file: tab3.tex
\begin{deluxetable}{lrrrr} 
\tabletypesize{\footnotesize}
\tablecolumns{5} 
\tablewidth{0pc} 
\tablecaption{Elemental abundances and comparison with 
literature.\label{tab:abund}}
\tablehead{ 
\colhead{Species} &
\colhead{This study} &
\colhead{Shetrone\tablenotemark{a}} &
\colhead{Sneden\tablenotemark{b}} &
\colhead{Cohen\tablenotemark{c}} 
}
\startdata
 \noalign{\vskip +0.5ex}
 \multicolumn{5}{c}{M 13 L598} \cr
 \noalign{\vskip  .8ex}%
 \hline
 \noalign{\vskip -2ex}\\
{\rm [Fe/H]} & $-$1.56 & $-$1.48 & $-$1.55 & \ldots \\
{\rm [O/Fe]} & 0.48 & 0.15 & 0.13 & \ldots \\
{\rm [Na/Fe]} & 0.06 & 0.08 & $-$0.03 & \ldots \\
{\rm [Mg/Fe]} & 0.22 & 0.23 & 0.28 & \ldots \\
{\rm [Al/Fe]} & 0.24 & 0.17 & 0.26 & \ldots \\
\cutinhead{M 13 L629}
{\rm [Fe/H]} & $-$1.63 & $-$1.46 & $-$1.62 & \ldots \\
{\rm [O/Fe]} & 0.40 & 0.25 & $-$0.13 & \ldots \\
{\rm [Na/Fe]} & 0.38 & $-$0.14 & 0.27 & \ldots \\
{\rm [Mg/Fe]} & 0.32 & 0.10 & 0.22 & \ldots \\
{\rm [Al/Fe]} & 0.74 & 0.53 & 0.70 & \ldots \\
\cutinhead{M 13 L70}
{\rm [Fe/H]} & $-$1.59 & $-$1.51 & $-$1.49 & $-$1.30 \\
{\rm [O/Fe]} & \ldots & $-$0.64 & $-$1.00 & $-$1.14 \\
{\rm [Na/Fe]} & 0.53 & 0.46 & 0.42 & 0.32 \\
{\rm [Mg/Fe]} & 0.07 & $-$0.16 & 0.02 & 0.29 \\
{\rm [Al/Fe]} & 1.27 & 1.11 & 1.16 & 0.64 \\
\cutinhead{M 13 L973}
{\rm [Fe/H]} & $-$1.61 & $-$1.45 & $-$1.57 & \ldots \\
{\rm [O/Fe]} & \ldots & $-$0.65 & $-$0.78 & \ldots \\
{\rm [Na/Fe]} & 0.44 & 0.30 & 0.45 & \ldots \\
{\rm [Mg/Fe]} & 0.04 & $-$0.26 & $-$0.08 & \ldots \\
{\rm [Al/Fe]} & 1.28 & 0.99 & 1.17 & \ldots \\
\cutinhead{M 71 A4}
{\rm [Fe/H]} & $-$0.91 & $-$0.76 & $-$0.76 & \ldots \\
{\rm [O/Fe]} & 0.65 & 0.21 & 0.37 & \ldots \\
{\rm [Na/Fe]} & 0.13 & 0.37 & 0.21 & \ldots \\
{\rm [Mg/Fe]} & 0.40 & 0.12 & \ldots & \ldots \\
{\rm [Al/Fe]} & 0.58 & 0.24 & \ldots & \ldots \\
\cutinhead{HD 141531}
{\rm [Fe/H]} & $-$1.72 & $-$1.67 & \ldots & \ldots \\
{\rm [O/Fe]} & 0.52 & 0.19 & \ldots & \ldots \\
{\rm [Na/Fe]} & $-$0.28 & $-$0.31 & \ldots & \ldots \\
{\rm [Mg/Fe]} & 0.45 & 0.36 & \ldots & \ldots \\
{\rm [Al/Fe]} & $-$0.01 & $-$0.06 & \ldots & \ldots \\
\enddata

\tablenotetext{a}{\citet{shetrone96a,shetrone96b}}
\tablenotetext{b}{M 13 giants were analyzed by \citet{sneden04a} 
for which [Fe/H] is the mean of [Fe/H]$_{\rm I}$ and 
[Fe/H]$_{\rm II}$. Na, Mg, and Al are referenced to [Fe/H]$_{\rm I}$ 
while O is referenced to [Fe/H]$_{\rm II}$. 
M 71 A4 was analyzed by \citet{sneden94}.}
\tablenotetext{c}{\citet{cohen05}}

\end{deluxetable}

%% file: tab4.tex
\begin{deluxetable}{lrrr} 
\tabletypesize{\footnotesize}
\tablecolumns{4} 
\tablewidth{0pc} 
\tablecaption{Abundance dependences on model parameters for
M 13 L629.\label{tab:parvar}}
\tablehead{ 
\colhead{Species} &
\colhead{\teff~+ 50} &
\colhead{$\log g$ + 0.2} &
\colhead{$\xi_t$ + 0.2}
}
\startdata
{\rm [Fe/H]}  & 0.01    & 0.01    & $-$0.05 \\
{\rm [O/Fe]}  & $-$0.01 & 0.04    & $-$0.01 \\
{\rm [Na/Fe]} & 0.04    & $-$0.10 & 0.06    \\
{\rm [Mg/Fe]} & 0.02    & $-$0.05 & 0.03    \\
{\rm [Al/Fe]} & 0.04    & $-$0.08 & 0.06    \\
\enddata

\end{deluxetable}

%% file: tab5.tex
\begin{deluxetable}{lcccccc} 
\tabletypesize{\footnotesize}
\tablecolumns{7} 
\tablewidth{0pc} 
\tablecaption{Magnesium isotope ratios for the program stars 
($^{24}$Mg:$^{25}$Mg:$^{26}$Mg).\label{tab:iso}}
\tablehead{ 
\colhead{Star} &
\colhead{5134.6\AA} &
\colhead{5138.7\AA} &
\colhead{5140.2\AA} &
\colhead{Final Ratio} &
\colhead{} &
\colhead{Literature} \\
\colhead{} &
\colhead{} &
\colhead{} &
\colhead{} &
\colhead{} &
\colhead{} &
\colhead{Value}
}
\startdata
M 13 L598 & 78:07:15 & 80:10:10 & 77:21:03 & 78:11:11 &  & 94:03:03\tablenotemark{a} \\
M 13 L629 & 69:10:22 & 70:08:22 & 72:20:08 & 70:11:19 &  & 70:15:15\tablenotemark{a} \\
M 13 L70  & 48:14:38 & 59:10:30 & 54:27:19 & 54:16:31 &  & 44:28:28\tablenotemark{a} \\
M 13 L973 & 44:15:41 & 50:05:45 & 50:23:27 & 48:13:39 &  & 50:25:25\tablenotemark{a} \\
M 71 A4   & 69:11:21 & 61:25:13 & 80:14:07 & 70:13:17 &  & \ldots \\
HD 141531 & 90:03:07 & 93:07:00 & 90:05:05 & 91:04:06 &  & 90:05:05\tablenotemark{a} \\
HD 141531 &          &          &          & 91:04:06 &  & 91:02:06\tablenotemark{b} \\
\enddata

\tablenotetext{a}{\citet{shetrone96b}}
\tablenotetext{b}{Y03}

\end{deluxetable}